\date{}
\documentclass[aps,pra,showpacs]{revtex4-1}
\usepackage{graphicx,psfrag,amsmath,amssymb,amsfonts,latexsym,color,dcolumn,bm}

\newcommand{\tG}{\tilde{G}}

\begin{document}
\title{Dissipation and decoherence effects on a moving particle in front of a dielectric plate}

\author{M. Bel\' en Far\'{\i}as $^1$~\footnote{mbelfarias@df.uba.ar}}
\author{Fernando C. Lombardo$^1$}

\affiliation{$^1$ Departamento de F\'\i sica {\it Juan Jos\'e
Giambiagi}, FCEyN UBA and IFIBA CONICET-UBA, Facultad de Ciencias Exactas y Naturales,
Ciudad Universitaria, Pabell\' on I, 1428 Buenos Aires, Argentina}

\date{today}
\begin{abstract}  
On this work, we consider a particle moving in front of a dielectric plate, and study two of the
most relevant effects of the vacuum field fluctuations: the dissipation, and the decoherence
of the particle's internal degrees of freedom. We consider the particle to follow a classical,
macroscopically-fixed trajectory. To study the dissipative effects, we calculate
the in-out effective action by functionally integrating over the vacuum field and the
microscopic degrees of freedom of both the plate and the particle. This in-out effective action 
develops an imaginary part, hence  a non-vanishing probability for the decay (because of friction)
of the initial vacuum state. We analyze how the dissipation is affected by the relative velocity between the particle and the plate
and the properties of the microscopic degrees of freedom.
In order to study the effects of decoherence over the internal degrees of freedom of the particle, we calculate the CTP
or Schwinger-Keldysh influence action, by functionally
integrating over the vacuum field and the microscopic degrees of freedom of the plate. We estimate the
decoherence time as the time needed by two different quantum configurations (of the internal degree of
 freedom of the particle) to be possible to differentiate from one another. We analyze the way in which
 the presence of the mirror affects the decoherence, and the possible ways to maximize or reduce its effects.

\end{abstract}
\maketitle
\section{Introduction}

Over the past few years, an increasing attention has been paid to the interaction between a particle and a (perfect or imperfect)
mirror or any dielectric surface \cite{others,dalvit,milonni,intravaia,pieplow,behuninhu,impens}. One of the main interests has been to
calculate the frictional force exerted over the particle by the plate, mediated by the vacuum field fluctuations. As in the case of
the quantum friction between two plates \cite{Pendry97,debate,vp2007}, there is still no general agreement about the nature of this frictional
force. These frictional effects are interesting, macroscopically observable consequences of the quantum nature of microscopic systems.
However, frictional and normal (Casimir) forces are not the only effects of the vacuum quantum fluctuations. Any quantum system that
interacts with an environment will suffer the process of decoherence, which is one of the main ingredients necessary to understand
the quantum-classical transition. The vacuum field is, clearly, an environment that cannot be switched off, since any particle (charged or
with non-vanishing dipole moment) will unavoidably interact with the electromagnetic field fluctuations. The effects of the electromagnetic
field over the coherence of the quantum state of a particle, and the way in which this effect is modified by the presence of a conducting plate, has already been
studied for interference experiments \cite{pazmazzi,maianeto}. However, in the many studies of the quantum friction over a moving particle
in front of a dielectric plate, the effects of decoherence have not yet been taken into consideration.  In this work, we will study the
decoherence of the internal degree of freedom of the particle. The loss of coherence of the particle's dipolar moment becomes relevant in any Ramsey interferometry experiment, where the depolarization of the atom could be macroscopically observed by means of the Ramsey fringes. In the case of a Rydberg atom, this phenomenon could be also observed as a decay of the Raby oscillations \cite{ramsey,rabi}.

On this work we consider a neutral particle moving in front of an imperfect mirror. The trajectory of the particle will be,
along this paper, kept as an externally-fixed variable. This accounts for many of the cases of interest, for example,
when the particle is the tip of a Atomic Force Microscope (AFM). When we specify the system, we will have the
particle moving at a constant velocity $v$, as is the most popular scenario in the literature \cite{others, dalvit}.
We are interested in the dynamics of the internal degree of freedom of the particle, that we will model as a 
quantum harmonic oscillator, being a simple model for the particle's electric dipole, and that will be coupled
in position to the vacuum field. We will also use a simple model
for the microscopic degrees of freedom of the mirror, as we have done in a previous work \cite{friction}: a set of
uncoupled harmonic oscillators, each of them also interacting locally with the vacuum field.  Even though this is a
simple model, it allows us to calculate some relevant quantities without much further assumptions.

We will study both the effects of friction and of decoherence over the moving particle, from the perspective of
quantum field theory. In order to do so, we will use two different but similar approaches. To study the friction,
we will follow the procedure presented in our previous work \cite{friction}, where we studied the friction generated
by the relative motion of two dielectric plates. We will calculate the imaginary part of the in-out effective action,
that will account for the dissipative effects. To study the decoherence, we will switch to the Closed Time Path (CTP) formalism and
calculate the action of influence of the environment (vacuum field + plate) over the particle, and use it to obtain
an estimation of the decoherence time, following an approach similar to the one in a previous work by some of
us \cite{Fosco:2007nz}.

The structure of this paper is as follows: in Section~\ref{sec:inout} we define the system we study,
and present the formalism in Minkowski space. We then calculate the in-out effective action for a specific
microscopic model, and obtain an expression for the imaginary part of the effective action (and hence the
dissipative effects over the system) as a function of the relative velocity. In Section \ref{sec:CTP} we
review the Schwinger-Keldysh formalism and obtain a general expression for the CTP influence action for the
internal degree of freedom of the particle, 
and use it to obtain the stochastic equations of motion for the particle's internal degree of freedom. In Section \ref{sec:deco}, 
we present a way to estimate of the decoherence time, and then analyze the way in which it is modified by the presence of the
plate. Finally, in Sec. \ref{sec:conc} we show our conclusions.

\section{In-out effective action and friction}\label{sec:inout}

\subsection{The system}\label{sec:sys}

\begin{figure}[h]
\centering
\includegraphics[scale=1]{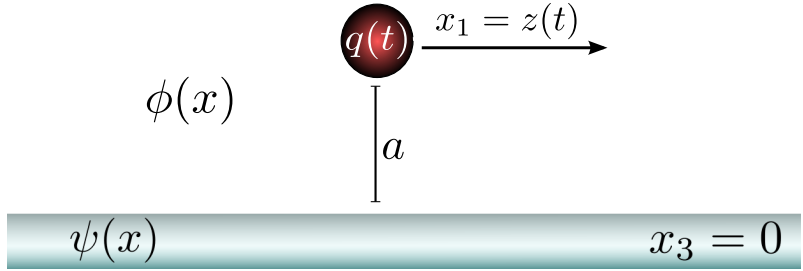}
\caption{\label{esquema} (Color online). A simple scheme of the system under consideration, where $\phi(x)$ is the vacuum field, $\psi(x)$ are the internal degrees of freedom of the plate, and $q(t)$ are the internal degrees of freedom of the particle, which follows a macroscopic trajectory $z(t)$ in the $x_1$ direction.}
\end{figure}

Let us consider a specific system, even though the formalism that is going to be developed among the next sections is general and could 
be used to study different problems. In the current work, we are interested in the dissipative effects over a particle that moves parallel to a 
flat, dielectric surface. The vacuum field shall be a non-massive scalar field $\phi(x)$ obeying the Klein-Gordon equation, interacting with both the 
particle and the internal degrees of freedom of the plate $\psi(x)$. The particle moves in a macroscopic, externally-fixed, uni-dimensional trajectory, 
in a plane parallel to the plate. The distance $a$ between the particle and the plate is also kept constant by an external source. We shall 
call $x_1$ the direction of movement of the particle, and $x_3$ the direction perpendicular to the plate. The particle also has an internal 
degree of freedom that we shall call $q$. A scheme of the system under consideration is shown in Fig. \ref{esquema}. We may write the classical action for the system as
\begin{equation}
S[\phi,\psi,q]=S_0^{\text{vac}}[\phi]+S_0^{\text{pl}}[\psi]+S_0^{\text{part}}[q]+S_{\text{int}}^{\text{pl}}[\phi,\psi]+S_{\text{int}}^{\text{part}}[\phi,q] \, ,
\end{equation}
where the Klein-Gordon action for the vacuum field, neglecting boundary terms, is given by
\begin{equation}
S_0^{\text{vac}}[\phi]=-\frac{1}{2} \int dx \phi(x) \left[ \partial_\mu \partial^\mu -i \epsilon \right] \phi(x) \, .
\end{equation}

The generating functional for the system will be given by
\begin{equation}
\mathcal{Z}=\int \mathcal{D}\phi \mathcal{D}\psi \mathcal{D}q e^{iS[\phi,\psi,q]} \, .
\end{equation}

The internal degrees of freedom of the plates can be integrated-out, resulting in an effective interaction potential $V(x,x')$ for the vacuum field. This procedure has already been performed in \cite{Fosco2011}, and the resulting classical action is
\begin{equation}
S[\phi,q]=S_{\text{eff}}[\phi]+S_0^{\text{part}}[q]+S_{\text{int}}^{\text{part}}[\phi,q] \, ,
\end{equation}
with
\begin{equation}
S_{\text{eff}}[\phi]=S_0[\phi]+\int dx dx' \phi(x) V(x,x') \phi(x') \, .
\end{equation}

The internal degree of freedom of the particle interacts with the vacuum field trough a current that we shall call $j(x)$. This current contains the information about the position and trajectory of the particle, and the strength of the coupling. The interaction term is, then,
\begin{equation}
S_{\text{int}}^{\text{part}}[\phi,q]=i \int dx \phi(x) j(x) \, .
\end{equation}
It is worth noticing that, at this point, the current $j(x)$ could make the vacuum field interact with any system of any given geometry: we have not yet specified that we are studying a moving, punctual particle. 

\subsection{The in-out effective action}\label{sec:in-out}

We would now like to obtain the effective action for the particle. That is, we are going to functionally integrate over the vacuum field, to obtain the generating functional
\begin{equation}
\mathcal{Z}[q]=\int \mathcal{D}\phi e^{i S[\phi,q]}=e^{i S_0[q]}\int \mathcal{D}\phi e^{i S_{\text{eff}}[\phi]+iS_{\text{int}}[\phi,q]} \, .
\end{equation}
This functional integral can be written as
\begin{align}
\label{eq:funcint}
\mathcal{Z}[q]=&\int \mathcal{D}\phi \exp\left[ -\frac{1}{2} \int dx dx' \phi(x) A(x,x') \phi(x') -\int dx \phi(x) j(x) \right] \nonumber \\
=& \left( \text{det} A \right)^{-\frac{1}{2}}
\exp\left[ \frac{1}{2} \int dx dx' j(x) A^{-1}(x,x') j(x')  \right] \, ,
\end{align}
where
\begin{equation}
A(x,x') = i \left[ \partial_\mu \partial^\mu - i \epsilon \right] \delta(x-x') - i V(x-x') \, .
\end{equation}

We need to obtain an operator $A^{-1}$ such that
\begin{equation}
A^{-1}(x,x')A(x,x')=A(x,x')=A^{-1}(x,x')=\delta(x-x') \, .
\end{equation}

In order to do so, we may write $A(x,x')=iA_0(x,x')+A_1(x,x')$, where
\begin{align*}
A_0(x,x')&=\delta(x-x')(\partial_\mu \partial^\mu - i \epsilon) \\
A_1(x,x')&= - i V(x,x') . 
\end{align*}
Now, the effective potential $V(x,x')$ is proportional to the coupling constant $\lambda$ between the vacuum field and the internal degrees of freedom of the plates. If this coupling is weak, then we can assume $A_1 << A_0$ and obtain $A^{-1}$ as an expansion in powers of $\lambda$. Keeping up to first order in $\lambda$, we write
\begin{equation}
\label{eq:Amenos1}
A^{-1}(x,x)\approx -i (\mathbb{I}+i A_0^{-1} A_1) A_0^{-1} \, . 
\end{equation}
And it is easy to prove that the operator from Eq. \eqref{eq:Amenos1} satisfies
\begin{equation}
A^{-1}A=A A^{-1} = \mathbb{I} + \mathcal{O}(\lambda ^2) \, . 
\end{equation}

Now, recalling the definition of $A_0$, it is clear that $A_0^{-1}$ is a Green function of the Klein-Gordon equation. We take it to be the Feynman propagator $A_0^{-1}(x,x')=G_F(x,x')$:
\begin{equation}
G_F(x,x')=\int \frac{dp}{(2\pi)^4}e^{-i p_\mu (x^\mu - x'^\mu)}\frac{1}{p_\mu p^\mu + i \epsilon} \, .
\end{equation}

The desired operator can be written as
\begin{equation}
A^{-1}(x,x')=-i\left(G_F(x,x')+ \int dy dy' G_F(x,y) V(y,y') G_F(y',x')\right) \, . 
\end{equation}

The only part of Eq. \eqref{eq:funcint} remaining to be calculated is the normalization factor $(\det A)^{1/2}$. However, this factor does 
not contribute to the connected diagrams: it involves only the interaction between the plate and the vacuum, and has no effect on the physics of the moving particle (or any other system we might be interested in, that could be introduced through the current $j(x)$). That being said, we might write the generating functional for the particle
\begin{equation}
\label{eq:genfunq}
\mathcal{Z}[q]=\exp \left\lbrace i S_0[q] -\frac{i}{2} \int dx dx' j(x) G_F(x,x')  j(x')- \frac{i}{2} \int dx dy dy' dx' j(x) G_F(x,y) V(y,y') G_F(y',x') j(x')  \right\rbrace \, .
\end{equation}

The effective action for the particle can be written as
\begin{equation}
\label{effact}
\Gamma[q]=S_0[q]+S_1[q]+S_2[q] \, ,
\end{equation}
where $S_1$ is the action that contains the influence of the vacuum over the particle, as if there was no dielectric plate, and is defined by
\begin{equation}
\label{eq:s1}
S_1[q]=-\frac{1}{2} \int dx dx' j(x) G_F(x,x')  j(x') \, .
\end{equation}
The term $S_2$ accounts for the influence of the plate over the particle, mediated by the vacuum field, and is defined by
\begin{equation}
S_2[q]=-\frac{1}{2} \int dx dy dy' dx' j(x) G_F(x,y) V(y,y') G_F(y',x')j(x') . 
\end{equation}

We might think of the effect of the plate on the particle as if mediated by a new effective propagator $\tilde{G}(x,x')$:
\begin{equation}
\label{Gtilde}
\tilde{G}(x,x')\equiv  \int dy dy'  G_F(x,y) V(y,y') G_F(y',x') \, ,
\end{equation}
and then $S_2$ can be written as
\begin{equation}
\label{eq:s2}
S_2[q]=-\frac{1}{2} \int dx  dx' j(x) \tilde{G}(x,x') j(x'). 
\end{equation}

But we are also interested in calculating the imaginary part of the in-out effective action for the whole system,
 since this quantity accounts for the dissipative effects over the system. The following calculations will be 
 completely analogous to the ones we have performed in \cite{friction} for the case of two plates in relative parallel motion.

To obtain the in-out effective action for the whole system, we need to integrate out the internal degree of freedom of 
the particle $q$, that is, we need to functionally integrate Eq. \eqref{eq:genfunq} over $q$:
\begin{equation}
\mathcal{Z}_{\text{sys}}=\int \mathcal{D}q \mathcal{Z}[q] . \end{equation}
But, instead of performing the functional integrations in this order (first over $\psi$, then over $\phi$, and last over $q$) we 
go back a few steps and integrate over $q$ before we integrate over the vacuum field $\phi$. The reason for doing this is
 purely of mathematical simplicity, and it does not affect the final expression for the effective action. We then obtain an 
 expression with two effective potentials, one accounting for the interaction of the plate with the vacuum field, and another 
 accounting for the interaction of the particle with the vacuum field. If we only take into consideration the terms involving both 
 the plate and the particle, then the effective action for small values of the coupling constants is \cite{friction}: 

\begin{equation}
\label{accionefectivamomentos}
\Gamma_I \approx \frac{-i}{2} \int \frac{dp}{(2\pi)^4} \frac{dq}{(2\pi)^4} G_F(p) V_{\text{plate}}(p,q) G_F (q)  V_{\text{particle}}(q,p) \, .
\end{equation}

\subsection{Specific model and results} \label{sec:inoutresults}

Now we would like to specify a concrete system of study. As we have already said in Sec. \ref{sec:sys}, even though, for the sake of clarity, through this work we have talked about a particle and a plate, this situation has not yet been specified. We have, so far, a complex, massless scalar field $\phi$ that interacts with another field $\psi$, that we call the internal degrees of freedom of a plate, but could be associated to any other system that we might be interested in. When we integrate those degrees of freedom out, we obtain a non-local effective potential $V(x,x')$ that contains the information about the characteristics of the other system. It could be a thin plate, a half-space, or any other geometry, with any kind of internal degrees of freedom. But if we do consider a thin, infinitesimal plate, occupying the $x_3=0$ plate, with internal degrees of freedom that interact locally in space with the vacuum field, then the effective potential will have the form: \cite{Fosco2011}
\begin{align}
V_{\text{plate}}(q,p)=&(2\pi)^3 \lambda(p_0)\delta(p_0-q_0)\delta^{(2)}(\bm{p}_\parallel-\bm{q}_\parallel)\, . 
\end{align}
And if the particle is considered to be punctual, moving along the $x_1$ axis with a constant velocity $v$, at a fixed distance $x_3=a$ above of the plate, and interacting locally in position with the vacuum field, then the effective potential results:
\begin{equation}
V_{\text{particle}}(q,p)=2\pi g(p_0-v p_1)\delta(p_0-q_0-v(p_1-q_1))e^{-ia(q_3-p_3)} \, . 
\end{equation}

These potentials take into account the geometry of the plate and the particle, but the information about their internal degrees of freedom and the nature of their interaction with the vacuum field are yet to be specified by the $\lambda$ and $g$ functions. The difference between a moving particle and a moving plate becomes clearer in position space, where the potential is localized in position by means of three Diracs' $\delta$-functions, one indicating that the particle is always on $x_3=a$ (this $\delta$-function was also present for the moving plate), other that the particle is always at $x_2=0$, and the third one indicating that it is always at $x_1=v t$.

By using the explicit shape of the potentials and integrating over $q_0$, $q_1$ and $q_2$, we find:
\begin{equation}
\Gamma_I \approx \frac{iT}{2(2\pi)^4} \int d^4p d_3 G(p_0,p_1,p_2,p_3) \lambda(p_0) G(p_0,p_1,p_2,q_3)  g(p_0-v p_1) e^{-i(q_3-p_3)a} \, ,
\end{equation}
where $T$ is total the time of flight of the particle. We consider the internal degrees of freedom of the plate to be uncoupled harmonic oscillators of frequency $\Omega$, each of them interacting locally in position with the vacuum field with a coupling constant $\lambda$; and the internal degree of freedom of the particle will also be a harmonic oscillator of frequency $\omega_0$, also interacting linearly and locally in position with the vacuum field, with a coupling constant $g$. This results \cite{friction} in
\begin{align}
\lambda(\omega)&=\frac{-\lambda^2}{\omega^2-\Omega^2+i\epsilon} ,\nonumber \\
g(\omega)&=\frac{-g^2}{\omega^2-\omega_0^2+i\epsilon} \, .
\end{align}

\begin{figure}[h]
\centering
\includegraphics[scale=1.5]{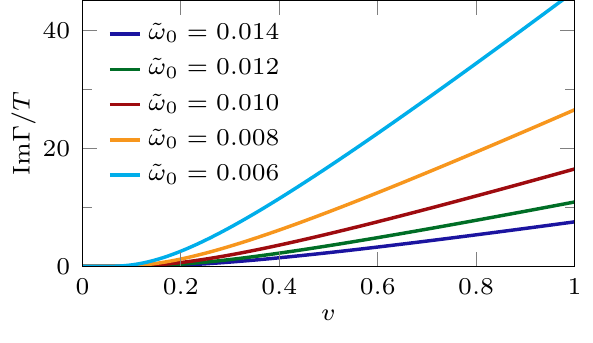}
\caption{\label{inout}(Color online). Imaginary part of the in-out effective action as a function of $v$, for $\tilde{\Omega}=\Omega a = 0.01$ and $\lambda=0.01$, in units of $g^2$.}
\end{figure}

The remaining integrations are performed in the exact same way as the integrations in Ref. \cite{friction}. The analytic result in $2+1$ dimensions is
\begin{equation}
\label{eq:gammainout}
{\mbox Im}\Gamma_I \approx \frac{T v \pi \lambda^2 g^2}{32 \tilde{\Omega} \tilde{\omega_0}} \frac{e^{-\frac{2}{v}\sqrt{(\tilde{\omega_0}+\tilde{\Omega})^2-v^2 \tilde{\Omega}^2}}}{(\tilde{\omega_0}+\tilde{\Omega})^2-v^2 \tilde{\Omega}^2} \, ,
\end{equation}
where $\tilde{\Omega}=\Omega a$ and $\tilde{\omega}_0=\omega_0 a$ are the dimensionless frequencies ($a$ is the distance between the particle and the plate). We show in Fig. \ref{inout} the imaginary part of the effective action as a function of the relative velocity $v$.

As we mentioned before, the imaginary part of the effective action implies the excitation of internal degrees of freedom on the mirror that influence the 
particle through the vacuum field. This signals a noncontact frictional effect. We can see that the dissipative effects are strongly suppressed as $v\rightarrow 0$.  This exponential vanishing of the dissipation effects has already been found, using different approaches, in previous works \cite{others,dalvit}. It is worth noticing that our coupling constant $g$ is the analog to the electric dipole moment $d$ appearing in the models used by other authors, since it accounts for the interaction between the particle's polarizability and the electromagnetic (vacuum) field. This means that the results presented here correspond to the $d^2$ contribution to the friction, and we will calculate the $d^4$ contribution within this approach in a future work. Lastly, let us recall that the $\lambda^2$ factor accounts for the interaction between the internal degrees of freedom of the plate, and this information is usually contained in the dielectric permitivity of the material, so there is no analog to our $\lambda$ factor appearing in the literature.

It is possible to see, from the general expression of Eq. \eqref{eq:gammainout}, that the case of resonance, for $\Omega = \omega_0$, coincides exactly with the expression of the imaginary part of the effective action per unit of area, for the case of two plates of frequency $\Omega$, studied in Ref. \cite{friction}. This is not that surprising, since in our model the plate and the particle/other plate only interact locally and the harmonic oscillators of the plate/s are uncoupled. In that work, we had set the coupling constant between the two plates to be equal, and that is the reason why our result was of order $\lambda^4$. It is important to notice that it does not mean that our result is to be compared with the $g^4$ results appearing in literature, since a factor $\lambda^2$ is to be considered part of the dielectric permitivity of the material.

\section{CTP action of influence and decoherence}\label{sec:CTP}

\subsection{The Schwinger-Keldysh or Close Time Path formalism}\label{sec:CTPform}

Up to this Section, we have considered the effective action for the system (the particle), and the influence action that mainly describes the dynamics of the particle after integration of the quantum fields. This in-out effective action cannot be applied in a straightforward way to the derivation of the equations of motion, since they would become neither real nor causal. As is well known, in order to get the correct effective equations of motion, one should compute the in-in, Schwinger-Keldysh or Closed Time Path Effective Action (CTPEA) \cite{CTP}, which also has information on the stochastic dynamics. The CTPEA is defined as:
\begin{equation}
\label{eq:defCTP}
e^{-i \Gamma_{\text{CTP}}}[q^+,q^-]=\int \mathcal{D}\phi^+ \mathcal{D} \phi^- e^{iS[q^+,\phi^+]-iS[q^-,\phi^-]}\equiv \int \mathcal{D}\phi e^{iS^\mathcal{C}[q,\phi]}
\end{equation}
where in the last step we have introduced the the CTP complex temporal path $\mathcal{C}$, going from minus to plus infinity $\mathcal{C}_+$ and backwards $\mathcal{C}_-$, with a decreasing (infinitesimal) imaginary part. Time integration over the contour $\mathcal{C}$ is defined by $\int_\mathcal{C} dt = \int_{\mathcal{C}_+} dt \int_{\mathcal{C}_-} dt$. The field in the last step is defined by $\phi(x,t)=\phi_{\pm}(x,t)$ if $t \in \mathcal{C}_\pm$. The equation above is useful because it has the structure of the usual in-out or the Euclidean effective
action, and we are able to use the results of the last section, but properly changing the contour of integration. This would lead to the appearance of the CTP propagators $G_{\alpha\beta}(\textbf{x},t;\textbf{x'},t')$, where $\alpha$ ($\beta$) $= \pm$ indicates that $t$ ($t'$) is on $\mathcal{C}_\pm$. For the free scalar field, the CTP propagators are:
\begin{align*}
G_{++}(x,x')=& \Delta_F(x,x') = -i \langle T \phi(x) \phi(x') \rangle \\
G_{--}(x,x')=& \Delta_D(x,x') = i \langle \hat{T} \phi(x) \phi(x') \rangle \\
G_{+-}(x,x')=& \Delta^-(x,x') = -i \langle  \phi(x') \phi(x) \rangle \\
G_{-+}(x,x')=& \Delta^+(x,x') = +i \langle  \phi(x) \phi(x') \rangle
\end{align*}

These CTP propagators satisfy the following properties, that will prove to be useful below: $(G_{++})^*=G_{--}$, $(G_{+-})^*=G_{-+}$, and $G_{++}-G_{--}=G_{+-}-G_{-+}$. 

We are now in a position to write the actions describing the influence of the environment on the particle within this formalism. We will do this in a general way, and later we will come back to our specific model. By performing the change in the contour of integration, $S_1$ results (see eq. \eqref{eq:s1}):
\begin{align}
\label{eq:s1if}
S_1^{\text{IF}}[q^+,q^-] &=-\frac{1}{2}\int dx dy \left[ j^+(x) G_{++}(x,y) j^+(y) - j^-(x) G_{--}(x,y) j^-(y) -j^+(x) G_{+-}(x,y) j^-(y) \right. \nonumber \\
&+\left. j^-(x) G_{-+}(x,y) j^+(y)\right] \, ,
\end{align}
where $j_\pm(x)$ is evaluated over $\mathcal{C}_\pm$. It is useful to define:
\begin{align*}
\Delta j(x)&=\frac{j^+(x)-j^-(x)}{2}\, ,\\
\Sigma j(x)&=\frac{j^+(x)+j^-(x)}{2}\, .\\
\end{align*}

By writing Eq. \eqref{eq:s1if} in terms of $\Delta j$ and $\Sigma j$, one obtains four different terms. However, by recalling the definition and properties of the different CTP propagators, it is easy to see that one of the terms vanishes, and two of the remaining terms are identical, with the result:
\begin{align*}
S_1^{\text{IF}}[q^+,q^-]&=-\frac{1}{2}\int dx dy \left\lbrace \Delta j (x) \left[ G_{++} (x,y) - G_{--} (x,y) - G_{-+}(x,y) + G_{+-}(x,y) \right] \Delta j (y) \right. \\
&+\left.  2 \Delta j (x) \left[ G_{++} (x,y) + G_{--} (x,y) + G_{-+}(x,y) + G_{+-}(x,y) \right] \Sigma j (y) \right\rbrace \, ,
\end{align*}
With some further considerations concerning the properties of the propagators mentioned above, it is possible to define the noise (diffusion) kernel
\begin{equation}
\label{eq:N1}
N_1(x,y)\equiv -i ( G_{++}(x,y)-G_{--}(x,y))= 2 \, \text{Im} \, G_{++}(x,y) \, ,
\end{equation}
associated with fluctuations, source of decoherence effects; and the dissipation kernel
\begin{equation}
\label{eq:D1}
D_1(x,y) \equiv  \frac{1}{2} \left[ G_{++}(x,y)+G_{--}(x,y)+G_{+-}(x,y)+G_{-+}(x,y)\right] = 2 \Theta (x_0-y_0) \text{Re} \, G_{++}(x-y) \, .
\end{equation}
Both kernels are real, and the dissipation kernel is explicitly causal. The influence action results, then 
\begin{equation}
\label{eq:S1CTP}
S_1^{\text{IF}}[q^+,q^-]=-\int dx dy \left[ i \Delta j (x)  N_1(x,y) \Delta j (y)  + 2 \Delta j (x) D_1(x,y) \Sigma j (y) \right] \, .
\end{equation}

This action accounts for the interaction of the particle with the vacuum, and all the propagators involved in the calculations correspond to free, scalar fields. Furthermore, in order 
to take the dielectric plate into consideration, we have to look at the $S_2$ term 
from Eq. \eqref{eq:s2}, which contains the effect of the plate over the particle up to first order in the coupling constant $\lambda$. By changing the contour of integration, we obtain an expression analogous to Eq. \eqref{eq:s1if}, but changing the free CTP propagator $G_{\alpha \beta}$ for an effective CTP propagator $\tilde{G}_{\alpha \beta}$, defined by:
\begin{align}
\label{eq:GtildeCTP}
\tilde{G}_{\alpha,\beta}(x,y)  \equiv &  \int dx' dy' \left[ G_{\alpha +}(x,x') V_{++}(x',y') G_{+\beta}(y',y) - G_{\alpha -}(x,x') V_{--}(x',y') G_{-\beta}(y',y) + \nonumber \right. \\
 & - \left. G_{\alpha +}(x,x') V_{+-}(x',y') G_{-\beta}(y',y) + G_{\alpha -}(x,x') V_{-+}(x',y') G_{+\beta}(y',y)\right]
 \end{align}
where $G_{\alpha \beta}$ is the free CTP propagator, and $V_{\alpha \beta}$ is the CTP expression for the effective potential. Both $G$ and $V$, being CTP Green functions, fulfill the relation $G_{++}-G_{--}=G_{+-}-G_{-+}$ and $V_{++}-V_{--}=V_{+-}-V_{-+}$, which allows us to prove that $\tilde{G}_{++}-\tilde{G}_{--}=\tilde{G}_{+-}-\tilde{G}_{-+}$, and thus:
\begin{equation}
\label{eq:S2CTP}
S_2^{\text{IF}}[q^+,q^-]=\int dx dy \left[ i \Delta j (x)  N_2(x,y) \Delta j (y)  + 2 \Delta j (x) D_2(x,y) \Sigma j (y) \right] \, ,
\end{equation}
with
\begin{equation}
\label{eq:N2}
N_2(x,y)\equiv \tG_{++}(x,y)+\tG_{--}(x,y)= 2 \text{Re} \, \tG_{++}(x,y)\, ,
\end{equation}
where we also used that $\tG_{++}=\left(\tG_{--}\right)^*$. The dissipation kernel
\begin{equation}
\label{eq:D2}
D_2(x,y) \equiv \frac{-i}{2} \left[ \tG_{++}(x,y)-\tG_{--}(x,y)-\tG_{+-}(x,y)+\tG_{-+}(x,y)\right]\, .
\end{equation}

On general grounds, the real and imaginary parts of the influence action can be associated with the dissipation and noise, respectively, and can be related by some integral equation known as the fluctuation-dissipation relation. As we shall see, the dissipation will be present in the generalized Langevin-like equation of motion for the 
internal degrees of freedom of the particle, and the noise kernel will be relevant in defining the correlation function for the noise source. 

\subsection{Our model}\label{sec:model}

We will now consider, as we have done in Sec. \ref{sec:inoutresults}, an infinitesimally thin plate occupying the plane $x_3=0$, formed by a set of uncoupled harmonic oscillators of frequency $\Omega$, each of them interacting locally in position with the vacuum field with a coupling constant $\lambda$. We have already found the CTP expression for the effective potential in a previous work: \cite{friction}
\begin{equation}
V^{\text{CTP}}(x-y)= \int \frac{dp}{(2\pi)^4} e^{-i p_\mu (x^\mu - y^\mu)} \lambda^2 \left( 
\begin{array}{c c}
\frac{1}{p_0^2-\Omega^2+i\epsilon} & \frac{i\pi \delta(p_0+\Omega)}{\Omega}  \\
\frac{i\pi \delta(p_0-\Omega)}{\Omega}  & \frac{1}{p_0^2-\Omega^2-i\epsilon}
\end{array}
\right) \, .
\end{equation}

The same kind of reasoning applies to the current $j(x)$. We now want it to describe a particle moving in the direction $x_1$, describing a trajectory $z(t)$, at a fixed distance $a$ from the plate. That is, its macroscopic coordinates are given by $x^\mu(t)=(z(t),0,a,t)$, and are kept fixed by an external source. On the other hand, we describe its internal degree of freedom as an uni-dimensional harmonic oscillator $q(t)$ of frequency $\omega_0$. This is the variable whose dynamics we now wish to study. So we have:
\begin{equation}
\label{eq:j}
j(x)=g q(t) \delta(x_1 - z(t)) \delta(x_2) \delta(x_3-a) \, ,
\end{equation}
where $g$ is the coupling constant between the vacuum and the internal degree of freedom of the particle. Now, as the classic trajectory $z(t)$ is macroscopic and externally fixed, we might assume that it remains the same in the different branches of the CTP integral. That is, we will from now on assume that $z^+(t)=z^-(t)$, where $z^\pm(t)$ is the classical function $z(t)$ integrated over the contour $\mathcal{C}_\pm$. This means that:
\begin{align*}
\Delta j (x) &= g \Delta q (t) \delta(x_1-z(t)) \delta(x_2) \delta(x_3) \\
\Sigma j (x) &= g \Sigma q (t) \delta(x_1-z(t)) \delta(x_2) \delta(x_3) .
\end{align*}

With these specific considerations, we can define, for $N \equiv N_1+N_2$:
\begin{equation}
N(t-t') \equiv g^2 \int d\textbf{x} d\textbf{x}' \delta(x_1-z(t)) \delta(x_2) \delta(x_3-a) N(x-x') \delta(x_1'-z(t')) \delta(x_2') \delta(x_3') \, ,
\end{equation}
and an analogous expression is obtained for $D\equiv D_1+ D_2$.

Within this model, we can write explicit expressions for $D_1$, $D_2$, $N_1$ and $N_2$ as integrals in momentum space. If the trajectory is $z(t)=v t$ for constant $v$, then:

\begin{equation}
\label{eq:d1}
D_1(t-t')= 2 g^2 \Theta(t-t') \text{Re} \int \frac{d^4 p}{(2\pi)^4} \frac{i e^{-i(p_0-v p_1)(t-t')}}{p_0^2-p_1^2-p_2^2-p_3^2+i\epsilon},
\end{equation}

\begin{equation}
\label{eq:n1}
N_1(t-t')= 2 g^2 \text{Im} \int \frac{d^4 p}{(2\pi)^4} \frac{i e^{-i(p_0-v p_1)(t-t')}}{p_0^2-p_1^2-p_2^2-p_3^2+i\epsilon},
\end{equation}

\begin{equation}
\label{eq:d2}
D_2(t-t')= \frac{\lambda^2 g^2}{2} \Theta(t-t') \text{Re} \int \frac{d^3 p}{(2\pi)^3} \frac{1}{p_0^2 - \Omega^2 +i \epsilon}\frac{e^{-i(p_0-v p_1)(t-t')+i i a \sqrt{p_0^2-p_1^2-p_2^2+i\epsilon}}}{p_0^2-p_1^2-p_2^2+i\epsilon},
\end{equation}

\begin{equation}
\label{eq:n2}
N_2(t-t')=  \frac{\lambda^2 g^2}{2}  \text{Im}  \int \frac{d^3 p}{(2\pi)^3} \frac{1}{p_0^2 - \Omega^2 +i \epsilon}\frac{e^{i(p_0-v p_1)(t-t')+2 i a \sqrt{p_0^2-p_1^2-p_2^2+i\epsilon}}}{p_0^2-p_1^2-p_2^2+i\epsilon}.
\end{equation}

Therefore,  the influence action for the moving particle in front of a dielectric plate is given, formally, by
\begin{equation}
\label{eq:effact}
S^{\text{IF}}[q^+,q^-]= \int dt dt'  \left[ i \Delta q (t)  N(t,t') \Delta q (t')  + 2 \Delta q (t) D(t,t') \Sigma q (t') \right] \, .
\end{equation}


\subsection{Real and causal equations of motion}

From Secs. \ref{sec:CTP} and \ref{sec:model}, it is easy to see that the (CTP) influence action has a real part, generated by the dissipation kernels $D_1$ and $D_2$, and an imaginary part, generated by the noise kernels $N_1$ and $N_2$. In order to obtain the real and causal equations of motion for $q(t)$, we must functionally derive $S[q^+,q^-]=S_0[q^+]-S_0^[q^-]+S^{\text{IF}}[q^+,q^-]$ with respect to $q^+$, and set $q^+=q^-=q$. However, this procedure would lead us to the equations for the mean value of $q(t)$, and the stochastic effect of the noise would not appear.

In order to see the influence of the noise on the equations of motion, it is necessary to consider a realization of the noise over the system, considering an stochastic source of noise $\xi(t)$. Following the well-known procedure for open quantum systems, we consider this stochastic force to have a Gaussian probability distribution given by:
\begin{equation}
\label{eq:Pxi}
P[\xi(t)]=N_\xi \exp \left\lbrace -\frac{1}{2} \int dt dt'\xi(t) N^{-1}(t,t') \xi(t') \right\rbrace \, . 
\end{equation}
It is easy to see that the inclusion of this stochastic source can be performed by adding to the generating functional the factor
\begin{equation}
\int \mathcal{D}\xi P[\xi] e^{-i \int dt \Delta q(t) \xi(t)}=e^{i \int dt dt' \Delta q(t) N(t,t') \Delta q(t')} \, ,
\end{equation}
which, as it is shown in the RHS of the last equation, leaves the generating functional unaltered. The modified influence action is, then
\begin{equation}
\hat{S}^{\text{IF}}[q^+,q^-,\xi]=2  \int dt dt'  \Delta q (t) D(t,t') \Sigma q (t')  - \int dt \Delta q(t) \xi(t) \, .
\end{equation}

And the equation of motion will be given by:
\begin{equation}
\left. \frac{\delta \left(S_0[q^+]-S_0[q^-]+\hat{S}^{\text{IF}}[q^+,q^-,\xi]\right)}{\delta q^+} \right|_{q^+=q^-=q}=0 \, .
\end{equation}

The result is:
\begin{equation}
\label{eq:eqmov}
\ddot{q}(t)+\omega_0^2 q(t) + \int dt' D(t,t') q(t') = \xi(t) \, ,
\end{equation}
where it is easy to see that the dissipation on the system comes from the kernel $D(t,t')$, and the fluctuations are generated by the stochastic force $\xi(t)$, that must fulfill, according to Eq. \eqref{eq:Pxi}:
\begin{align*}
\langle \xi(t) \rangle &= 0 \\
\langle \xi(t) \xi(t') \rangle &= N(t,t')
\end{align*}

This is a generalized Langevin equation, with classical noise $\xi$ and dissipation, that satisfies the fluctuation-dissipation theorem.  Each part of the environment that we include leads to a further \textit{dissipative} term on the left-hand side of Eq.(\ref{eq:eqmov}) (i.e. 
$D_1$ and $D_2$ as we named before) with a countervailing noise term on the right-hand side. Of course it is very difficult to
solve this stochastic equation analytically. It is difficult to imagine an \textit{ab initio} derivation of the dissipative and noise terms from the full quantum theory. In this sense, a reasonable alternative is to analyze phenomenological stochastic equations numerically and check the robustness of the predictions against different choices of the dissipative kernels and of the type of noise. 

\section{Decoherence of the particle's internal degrees of freedom}\label{sec:deco}

The absence of quantum interference between the stationary phase solutions to the classical stochastic equations, $q(t)$, is manifested through the increasing	decay of the non-diagonal terms of	the reduced density matriz $\rho[q^+,q^-, t]$.	This	leads to the crucial notion of a decoherence time $t_D$, after which	 $\rho$	(or,	more	 exactly,	its	real	part)	is effectively diagonal. Eq. (\ref{eq:eqmov}) is the classical stochastic equation that we are looking for but, as it stands, is only guaranteed to describe classical 
degrees of freedom $q_{\rm class}(t)$ after decoherence (see in another context Refs. \cite{PLBRay,NPB}).

The notion of consistent histories provides an alternative approach to classicality, opposed to trying to solve the master equation, and it is normally used in open quantum systems.  Quantum evolution can be considered as a coherent superposition of fine-grained histories. Since we need to be able to distinguish different classical system configurations evolving in time, we work in the basis of amplitudes $q(t)$. If one defines the c-number	$q(t)$ as specifying a fine-grained history, the quantum amplitude for that history is 
$\Psi \sim e^{S[q]}$  (we work in units in which $\hbar = 1$) \cite{PLB07}.
In the quantum open system approach that we have adopted here, we are concerned with coarse-grained histories 

\begin{equation}
\Psi[\alpha] = \int {\cal D}q e^{S[q]} \alpha[q],
\end{equation}
where $\alpha[q]$ is the filter function that defines the coarse-graining. In the first instance this filtering corresponds to tracing over all the degrees of freedom of the 
composite environment. From this we define the decoherence functional for two coarse-grained histories as

\begin{equation}
{\cal D}[\alpha^+,\alpha^-] = \int {\cal D}q^+  {\cal D}q^- e^{i (S[q^+] - S[q^-])} \alpha^+[q^+] \alpha^-[q^-].
\end{equation}

${\cal D}[\alpha^+,\alpha^?]$ does not factorize because the histories $q^\pm$ are not independent. Decoherence means physically that the different coarse-graining histories making up the full quantum evolution acquire individual reality, and may therefore be assigned definite probabilities in the classical sense. A necessary and sufficient condition for the validity of the sum rules of probability theory (i.e. no quantum interference
terms) is \cite{Griffiths}

\begin{equation}
{\rm Re}{\cal D}[\alpha^+,\alpha^-] \approx 0,\end{equation}
when $\alpha^+ \not= \alpha^-$, (although in most cases the stronger condition ${\cal D} [\alpha^+,\alpha^-] \approx 0$ holds) \cite{Ommes}. Such histories are consistent 
\cite{GellMann}.

For our particular application, we wish to consider as a single coarse-grained history all those fine-grained ones where the solution $q(t)$ remains 
close to a prescribed classical configuration $q_{\rm cl}$. The filter function takes the form $\alpha_{\rm cl} [q]= \int {\cal D}J e^{i \int J (q - q_{\rm cl} } \alpha_{\rm cl} [J]$.Using 
that $ J q = \int dx J(x) q[x]$,  we may write the decoherence functional between two classical
histories in terms of the closed-path-time generating functional. In principle, we can examine adjacent general classical solutions for their consistency but, in practice, it is simpler to restrict ourselves to particular solutions $q^{\pm}$, according to the nature of the decoherence that we are studying. All in all, the decoherence functional results in 

\begin{equation}
{\cal D} [q_{\rm cl}^+,q_{\rm cl}^-] \approx F [q_{\rm cl}^+,q_{\rm cl}^-],
\end{equation}
where $F [q_{\rm cl}^+,q_{\rm cl}^-]$ is the	Feynman-Vernon influence	functional
(IF) \cite{vernon}. The influence functional is written in terms of the influence action $S^{\rm IF}[q^+, q^-]$ as $F = \exp[- i S^{\rm IF}]$. As a result,

\begin{equation}
\vert {\cal D}  [q_{\rm cl}^+,q_{\rm cl}^-] \vert \sim \exp\{ i {\rm Im} S^{\rm IF} [q_{\rm cl}^+,q_{\rm cl}^-]\} ,
\end{equation}
where $S^{\rm IF}$ is the contribution to the action due to the, in our case, composite environment.
From this viewpoint, once we have chosen the classical solutions of interest, adjacent histories become consistent at the time $t_D$, for which
$1 \approx  {\rm Im} S^{\rm IF}\vert_{t=t_D}$ \cite{PLB07}. This decoherence time will become relevant in any Ramsey interferometry experiment \cite{ramsey}. For times larger than $t_D$, the Ramsey fringes will no longer be distinguishable, expressing the depolarization of the atom.

\subsection{Imaginary part of $S^{\text{IF}}_1$}

Let us recall the expression for $N_1(t,t')$ given by Eq. \eqref{eq:N1}, and consider two classical trajectories $q_{\rm cl}(t)$ with different amplitudes but the same frequency.
\begin{equation}
\Delta q_{\rm cl} (t) = \Delta q_0 \cos(\omega_0 t).
\end{equation}

So we can write an expression for the imaginary part of the influence action for the particle in presence of the vacuum field (ignoring the plate), as:
\begin{align*}
\text{Im} S^{\text{IF}}_1=\frac{-g^2 \Delta q_0^2}{4} \text{Im} \int \frac{dp_0 d\textbf{p}}{(2\pi)^4} \frac{1}{p_0^2 - \textbf{p}^2 + i \epsilon} &\left\lbrace \int dt dt' e^{i (p_0 - v p_1 + \omega_0)(t-t') } + \int dt dt' e^{i (p_0 - v p_1 - \omega_0)(t-t') } + \right. \\
 &+  \left. \int dt dt' e^{i (p_0 - v p_1 + \omega_0)(t+t')} +\int dt dt' e^{i (p_0 - v p_1 - \omega_0)(t+t') }\right\rbrace. 
\end{align*}
All the temporal integrations result in Dirac delta functions. The last two terms vanish because the resulting deltas cannot be fulfilled simultaneously. From each non-vanishing term, an infinite $\delta(0)$ is obtaining, accounting for the total time of integration $T$ (time of flight of the particle). Thus the integrals over $p_0$ can trivially be performed, and the remaining terms are:
\begin{equation}
\text{Im} S^{\text{IF}}_1=\frac{-g^2 \Delta q_0^2 T}{4} \text{Im} \int \frac{dp_0 d\textbf{p}}{(2\pi)^2} \frac{\delta(p_0+v p_1 - \omega_9) \delta(p_0+v p_1 + \omega_0)}{p_0^2 - \textbf{p}^2 + i \epsilon}. 
\end{equation}
Now, if we take the limit $\epsilon \rightarrow 0$, then:
\begin{equation}
\frac{1}{{p_0^2 - p_\parallel^2 + i \epsilon}} \rightarrow \text{p.v} \left( \frac{1}{p_0^2 - \textbf{p}^2} \right) - i \pi \delta\left(p_0^2 - \textbf{p}^2\right)
\end{equation}
where ${\rm p.v}$ denotes the Cauchy principal value. So we have:
\begin{equation}
\text{Im} S^{\text{IF}}_1=\frac{-g^2 \Delta q_0^2 T}{4} \int \frac{dp_0 d\textbf{p}}{(2\pi)^2}\left[ \delta\left(p_0+v p_1 - \omega_0\right)+ \delta\left(p_0+v p_1 + \omega_0\right) \right]\delta \left(p_0^2-\textbf{p}^2\right).
\end{equation}

Let us now consider the case in $2+1$ dimensions, setting $p_2 \equiv 0$ ($x_2$ is the direction with translational symmetry), then, integrating over $p_0$, we have:
\begin{equation}
\text{Im} S^{\text{IF}}_1=\frac{ \pi g^2 \Delta q_0^2 T}{4}  \int \frac{dp_1 dp_3}{2\pi}  \left[\delta\left((-v p_1 + \omega_0)^2 -(p_1^2+p_3^2)\right)+ \delta\left((v p_1 + \omega_0)^2 -(p_1^2+p_3^2)\right) \right], 
\end{equation}
and, assuming the $\delta$-functions as functions of $p_3$, we have four terms:
\begin{equation}
\text{Im} S^{\text{IF}}_1=\frac{ \pi g^2 \Delta q_0^2 T}{4}  \int \frac{dp_1 dp_3}{2\pi}  \left[\frac{\delta\left( p_3 - \sqrt{(v^2-1)p_1^2 \pm 2 \omega_0 v p_1 + \omega_0^2}\right)}{2\sqrt{(v^2-1)p_1^2 \pm 2 \omega_0 v p_1 + \omega_0^2}}+ \frac{\delta\left( p_3 + \sqrt{(v^2-1)p_1^2 \pm 2 \omega_0 v p_1 + \omega_0^2}\right)}{2\sqrt{(v^2-1)p_1^2 \pm 2 \omega_0 v p_1 + \omega_0^2}} \right].
\end{equation}

Since the integration is over the real values of $p_3$, the integral of these $\delta$-functions can only be non-vanishing if the squared root is a real number. It is easy to see, by looking carefully at the argument of the squared root, that the "$+$" terms can only be non-zero if:
\begin{equation}
\frac{-\omega_0}{1+v}<p_1<\frac{\omega_0}{1-v}
\end{equation}
and the "$-$" terms are non-vanishing only if:
\begin{equation}
\frac{-\omega_0}{1-v}<p_1<\frac{\omega_0}{1+v} . 
\end{equation}
So that we can write the remaining integral over a bounded set of $p_1$:
\begin{equation}
\text{Im} S^{\text{IF}}_1=\frac{ g^2 \Delta q_0^2 T}{32 \pi}  \left[\int_{-\frac{\omega_0}{1+v}}^{\frac{\omega_0}{1-v}}dp_1\frac{1}{\sqrt{(v^2-1)p_1^2 + 2 \omega_0 v p_1 + \omega_0^2}}+ \int_{-\frac{\omega_0}{1-v}}^{\frac{\omega_0}{1+v}}dp_1\frac{1}{\sqrt{(v^2-1)p_1^2 - 2 \omega_0 v p_1 + \omega_0^2}} \right]. 
\end{equation}

Now the integrands are of the form $(- a (p_1\pm b)^2 )^{-1/2}+ c$, with  $a= (1-v^2)$, $b=\frac{\omega_0 v}{1-v^2}$ and $c= \frac{\omega_0^2}{1-v^2}$. 
%
Thus, we have to take carefully the limit  when $p_1$ tends to the integration boundaries, and we find that the result of each integral is $\frac{\pi}{\sqrt{1-v^2}}$, which means that:
\begin{equation}
\label{eq:s12d}
\text{Im} S^{\text{IF}}_1=\frac{ g^2 \Delta q_0^2 T}{16 \sqrt{1-v^2}}. 
\end{equation}

\subsection{Imaginary part of $S^{\text{IF}}_2$}

Let us recall the expression for $N_2(t,t')$ given by Eq. \eqref{eq:N2}, and consider two classical trajectories $q(t)$ with different amplitudes but the same frequency.
\begin{equation}
\label{eq:deltaq}
\Delta q_{\rm cl} (t) = \Delta q_0 \cos(\omega_0 t)
\end{equation}
By inserting these results in Eq. \eqref{effact}, we are able to write an expression for the imaginary part of the action of influence, given by:
\begin{align*}
\text{Im}S^{\text{IF}}_2= \frac{g^2 \lambda^2 \pi^2 \Delta q_0^2}{2} \text{Im} \int \frac{dp_0 dp_\parallel}{(2\pi)^3} \frac{e^{2i a \sqrt{p_0^2 - p_\parallel^2 + i \epsilon}}}{(p_0^2 - p_\parallel^2 + i \epsilon)(p_0^2 - \Omega^2 + i \epsilon)} &\left\lbrace \int dt dt' e^{i (p_0 - v p_1 + \omega_0)(t-t') } + \int dt dt' e^{i (p_0 - v p_1 - \omega_0)(t-t') } + \right. \\
 &+  \left. \int dt dt' e^{i (p_0 - v p_1 + \omega_0)(t+t')} +\int dt dt' e^{i (p_0 - v p_1 - \omega_0)(t+t') }\right\rbrace
\end{align*}

%
Following the same procedure than before, we find, 
\begin{equation}
\label{eq:dosterminos}
\text{Im}S^{\text{IF}}_2= \frac{g^2 \pi^2 \Delta q_0^2 \lambda^2 T}{2} \text{Im} \int \frac{ dp_\parallel}{(2\pi)^2} \frac{e^{2i a \sqrt{(v p_1 - \omega_0)^2 - p_\parallel^2 + i \epsilon}}}{((v p_1 - \omega_0)^2 - p_\parallel^2 + i \epsilon)((v p_1 - \omega_0)^2 - \Omega^2 + i \epsilon)} + \omega_0 \leftrightarrow -\omega_0 .
\end{equation}

We will now consider, again, the case of $2+1$ dimensions, setting $p_2 = 0$. By changing variables $k \equiv v p_1$, we have:
\begin{equation}
\text{Im}S^{\text{IF}}_2= \frac{g^2 \pi^2 \Delta q_0^2 \lambda^2 T}{2} \text{Im} \int \frac{ dk}{(2\pi)} \frac{e^{\frac{2i a}{v} \sqrt{P(k) + i \epsilon}}}{(P(k) + i \epsilon)(A(k) + i \epsilon)} + \omega_0 \leftrightarrow -\omega_0 \, . 
\end{equation}
Here $P(k)=(k - \omega_0)^2 v^2 - k^2$ and $A(k)=(k-\omega_0)^2-\Omega^2$. We will begin by considering the first term. If $P(k)$ had a definite sign, we could get rid of the $i \epsilon$ accompanying it. Since $P(k)$ has two real roots, located at 
$k_+ = \frac{v \omega_0}{1+v}$ and
$k_- = -\frac{ v \omega_0}{1-v}$
 we will consider three different regions:
 \begin{align*}
 &\text{(I)} \rightarrow \quad \frac{v \omega_0}{1+v} < k & P(k)<0 \\
 &\text{(II)} \rightarrow \quad -\frac{ v \omega_0}{1-v}  k < \frac{v \omega_0}{1+v}  & P(k)>0 \\
 &\text{(III)} \rightarrow \quad k < -\frac{ v \omega_0}{1-v} & P(k)<0
 \end{align*}

Now, within each region, $P(k)$ has a definite sign, and the limit $\epsilon \rightarrow 0$ can be taken on the terms involoving $P(k)$.

On the other hand, the integrand will have two poles associated with the zeros of $A(k)$, which are located at $k_1= \omega_0 + \Omega$ and $k_2 = \omega_0 - \Omega$. It is easy to see that $k_1$ is always located in region (III), and that $k_2$ could be on any region, depending on the values of the external parameters of the problem: $v$, $\omega_0$ and $\Omega$. With a little algebra, we find that $k_2$ is on region (II) if and only if:
\begin{equation}
\label{eq:condition}
\frac{\omega_0}{1+v}< \Omega < \frac{\omega_0}{1-v} \, .
\end{equation} 
This condition is important in order to explicitly calculate the integrals.

Let us start computing the integral over region (II). In order to do so, we will have in mind that, when ${\epsilon \rightarrow 0} $:
\begin{equation}
\frac{1}{A(k)+i \epsilon} \rightarrow \text{p.v.} \left(\frac{1}{A(k)}\right) - i \pi \delta \left( A(k) \right)
\end{equation}
and, since $P(k)>0$ on this region, we may write:
\begin{equation*}
\exp\left(\frac{2ia}{v}\sqrt{P(k)}\right) = \cos \left(\frac{2ia}{v}\sqrt{P(k)}\right) + i \sin \left(\frac{2ia}{v}\sqrt{P(k)}\right) \, .
\end{equation*}

If the condition \eqref{eq:condition} is fulfilled, then $A(k)$ will have a zero in the integration region, and $\delta(A(k))$ will be nonzero. We will have two terms contributing to the imaginary part of the effective action. The first one will be:
\begin{equation}
\int_{k_-}^{k^+}  \frac{dk}{2 \pi} (-v \pi) \frac{\cos\left[\frac{2a}{v}\sqrt{P(k)}\right]}{P(k)} \delta\left( A(k) \right) = - \frac{v}{4 \Omega} \frac{\cos\left[\frac{2a}{v}\sqrt{\Omega^2 v^2 - (\omega_0 - \Omega)^2}\right]}{\Omega^2 v^2 - (\omega_0 - \Omega)^2}
\end{equation}
where we have inserted the explicit expressions for $P(k_2)$ and $|A'(k_2)|$. The other term is:
\begin{equation}
\frac{v}{2\pi} \int_{k_-}^{k^+} dk \frac{\sin \left(\frac{2ia}{v}\sqrt{P(k)}\right)}{P(k)} \text{p.v.} \left(\frac{1}{A(k)}\right)
\end{equation}
Now, if the condition \eqref{eq:condition} is not satisfied, then there the integrand has no poles on the integration region, and we can take $\epsilon \rightarrow 0$ everywhere. This results in one term contributing to the imaginary part of the effective action:
\begin{equation}
\frac{v}{2\pi} \int_{k_-}^{k^+} dk \frac{\sin \left(\frac{2ia}{v}\sqrt{P(k)}\right)}{P(k)A(k)}
\end{equation}

Let us now compute the integral over regions (I) and (III). Since on this region $P(k)<0$, we may write:
\begin{equation}
\text{Im} \left\lbrace \left[ \int_{-\infty}^{k_-} dk + \int_{k_+}^{+\infty}\right]\underbrace{\frac{e^{-\frac{2a}{v}\sqrt{-P(k)}}}{P(k)}}_{\in \mathbb{R}}\right\rbrace v \frac{1}{A(k)+i \epsilon}
\end{equation}
and, when we take $\epsilon \rightarrow 0$, the only contributions to the imaginary part will come from the $- i \pi \delta$ term. The only remaining question is how many zeros does $A(k)$ have on the integration region. As we already said, $k_1$ is always on region (III), so we will always have a contribution coming from $k_1$, that is:
\begin{equation}
\frac{v}{4 \Omega} \frac{e^{-\frac{2a}{v}\sqrt{(\omega_0+\Omega)^2-\Omega^2 v^2}}}{(\omega_0+\Omega)^2-\Omega^2 v^2}
\end{equation}
And, when the condition \eqref{eq:condition} is not fulfilled, then $k_2$ is also on region (I) or on region (III), so we will have an extra contribution of the form:
\begin{equation}
\frac{v}{4 \Omega} \frac{e^{-\frac{2a}{v}\sqrt{(\omega_0-\Omega)^2-\Omega^2 v^2}}}{(\omega_0-\Omega)^2-\Omega^2 v^2}
\end{equation}

So far, we have computed the result of the first term on Eq. \eqref{eq:dosterminos}. The procedure for computing the second term is completely analogous, so we are now in a position to write the result for the imaginary part of the action of influence. We define:
\begin{align}
\label{eq:cals1}
\mathcal{S}_1=&\frac{1}{2\pi} \int_{-\frac{v \omega_0}{1-v}}^{\frac{v \omega_0}{1+v}} dk \frac{\sin[\frac{2a}{v} \sqrt{(k-\omega_0)^2v^2-k^2}]}{((k-\omega_0)^2v^2-k^2)((k-\omega_0)^2-\Omega^2)} + \frac{1}{2\pi} \int_{-\frac{v \omega_0}{1+v}}^{\frac{v \omega_0}{1-v}} dk \frac{\sin[\frac{2a}{v} \sqrt{(k+\omega_0)^2v^2-k^2}]}{((k+\omega_0)^2v^2-k^2)((k+\omega_0)^2-\Omega^2)} \nonumber \\
+& \frac{1}{2\Omega} \frac{e^{\frac{2a}{v} \sqrt{(\omega_0+\Omega)^2-v^2\Omega^2}}}{(\omega_0+\Omega)^2-v^2\Omega^2} + \frac{1}{2\Omega} \frac{e^{\frac{2a}{v} \sqrt{(\omega_0-\Omega)^2-v^2\Omega^2}}}{(\omega_0-\Omega)^2-v^2\Omega^2}
\end{align}

and

\begin{align}
\label{eq:cals2}
\mathcal{S}_2&=\frac{1}{2\pi} \int_{-\frac{v \omega_0}{1-v}}^{\frac{v \omega_0}{1+v}} dk \frac{\sin[\frac{2a}{v} \sqrt{(k-\omega_0)^2v^2-k^2}]}{(k-\omega_0)^2v^2-k^2} \text{p.v} \left( \frac{1}{(k-\omega_0)^2-\Omega^2} \right) - \frac{1}{2\Omega} \frac{\cos[\frac{2a}{v} \sqrt{v^2\Omega^2-(\omega_0-\Omega)^2}]}{v^2\Omega^2-(\omega_0-\Omega)^2} \nonumber \\
&+ \frac{1}{2\pi} \int_{-\frac{v \omega_0}{1+v}}^{\frac{v \omega_0}{1-v}} dk \frac{\sin[\frac{2a}{v} \sqrt{(k+\omega_0)^2v^2-k^2}]}{(k+\omega_0)^2v^2-k^2} \text{p.v} \left( \frac{1}{(k-\omega_0)^2+\Omega^2} \right)+  \frac{1}{2\Omega} \frac{e^{\frac{2a}{v} \sqrt{(\omega_0+\Omega)^2-v^2\Omega^2}}}{(\omega_0+\Omega)^2-v^2\Omega^2}  \, ,
\end{align}
{where the remaining integrals over $k$ can be performed numerically.} The imaginary part of the action of influence is then given by
\begin{equation}
\text{Im} S^{\text{IF}}_2 = \frac{g^2 \pi^2 \Delta q_0^2 \lambda^2 T v}{2} \left\lbrace \begin{array}{c c c} 
\mathcal{S}_1 & \text{if} & \Omega<-\frac{\omega_0}{1+v} \, \text{or} \, \frac{\omega_0}{1+v}<\Omega \\
\mathcal{S}_2 & \text{if} & -\frac{\omega_0}{1+v}<\Omega < \frac{\omega_0}{1+v}
\end{array} \right. \, .
\end{equation}

\subsection{Estimation of the decoherence time}

We are now in a position to estimate the decoherence time for the particle. The total imaginary part of the action of influence (up to second order in $\lambda$) is:
\begin{equation}
\text{Im} S^{\text{IF}} = \frac{ g^2 \Delta q_0^2 T }{2}\left( \frac{ 1}{8 \sqrt{1-v^2}} + \lambda^2 v
\left\lbrace \begin{array}{c c c} 
\mathcal{S}_1 & \text{if} & \Omega<-\frac{\omega_0}{1+v} \, \text{or} \, \frac{\omega_0}{1+v}<\Omega \\
\mathcal{S}_2 & \text{if} & -\frac{\omega_0}{1+v}<\Omega < \frac{\omega_0}{1+v}
\end{array} \right. \right)
\end{equation}

\begin{figure}
\includegraphics[scale=1.5]{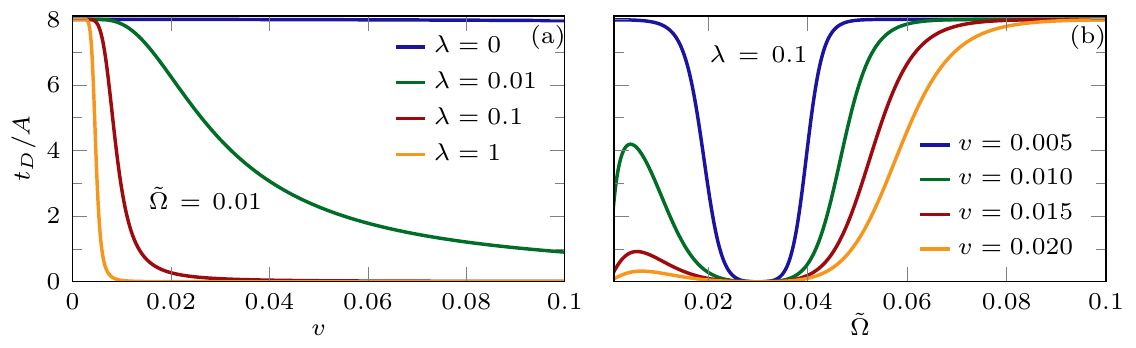} 
\caption{\label{tD}Estimation of the decoherence time, in units of a global factor $A=\frac{g^2 \Delta q_0^2}{2}$, as a function of \textbf{(a)} the relative velocity $v$ and \textbf{(b)} the dimensionless material frequency $\tilde{\Omega}$, for $\tilde{\omega}_0 = 0.03$.}
\end{figure}

Now, as we have detailed on the beginning of this Section, the decoherence time can be estimated by the time of flight of the particle when $\text{Im} S^{\text{IF}} \sim 1$, so that:
\begin{equation}
\label{eq:time}
t_D \sim \frac{ 2 }{g^2 \Delta q_0^2} \frac{1}{\left( \frac{ 1}{8 \sqrt{1-v^2}} + \lambda^2 v
\left\lbrace \begin{array}{c c c} 
\mathcal{S}_1 & \text{if} & \Omega<-\frac{\omega_0}{1+v} \, \text{or} \, \frac{\omega_0}{1+v}<\Omega \\
\mathcal{S}_2 & \text{if} & -\frac{\omega_0}{1+v}<\Omega < \frac{\omega_0}{1+v}
\end{array} \right. \right)}
\end{equation}

where $\mathcal{S}_1$ and $\mathcal{S}_2$ are defined in Eq. \eqref{eq:cals1} and \eqref{eq:cals2} respectively.
{After numerically performing these integrals, we are in a position to plot, in Fig. \ref{tD} \textbf{(a)}, the estimation of the decoherence time as a function of the velocity of the particle, for different values of the coupling constant between the plate and the vacuum, $\lambda$.} There is a global factor $A=\frac{ 2 }{g^2 \Delta q_0^2}$, that shows that the decoherence time is reduced for larger values of the coupling constant between the particle and the vacuum field, and for larger difference in the amplitudes of the classical trajectories under consideration.

Now, as can be seen in Fig. \ref{tD} \textbf{(a)}, the presence of the plate reduces the decoherence time, but only for non-vanishing relative velocity. For a particle that is at rest or moving at very small velocities with respect to the plate, the decoherence time is not reduced, even for greater values of the coupling constant between the plate and the vacuum field (and hence stronger interaction between the particle and the plate, mediated by the vacuum field). Now, for non-vanishing velocities, the decoherence of the particle is greater the stronger the coupling between the plate and the vacuum field.

In Fig. \ref{tD} \textbf{(b)}, $t_D$ is shown as a function of the plate's characteristic dimensionless frequency $\tilde{\Omega}$, for different values of its macroscopic velocity $v$. In the graphic, a clear minimum appears for every value of $v$, and it is located in $\tilde{\Omega}=0.03=\tilde{\omega}_0$. This means that the decoherence is maximal in the resonant case, hence making the decoherece time vanish. Far from the resonance, that is, for values of $\Omega>>\omega_0$ (or the opposite limit when possible), the decoherence time tends to the limiting value that corresponds to the case $\lambda=0$ (the case with no plate).

From the results shown above, we can see that, in this simple model, the presence of the plate enhances the decoherence over the particle. {The decoherence effects can be maximized by an appropiate choice of the atom's fine-grained history ($\Delta q_0$ the classical amplitud difference in our example)  and the plate's material (for a relation between the $\lambda(\omega)$ function of the material and its dielectric permitivity $\epsilon(\omega)$, see Ref. \cite{neumann}). As it has already been discussed \cite{pazmazzi}, the decoherence effects given solely by the vacuum field are neglectable. The coupling between any neutral atom and the vacuum field is very weak, resulting in an almost infinite decoherence time. However, from our results it can be seen that, by carefully choosing the plate's material (i.e., close to the resonance in Fig. \ref{tD} \textbf{(b)} ), the decoherence time could be drastically reduced. We also show that the relative velocity contributes to the enhancing of the decoherence effects.}

\section{Conclusions}\label{sec:conc}

In this paper we have used a functional approach to study the effects of quantum vacuum
fluctuations on a particle moving parallel to an imperfect mirror. We have presented a simple model in which
the vacuum field is a massless, real, scalar field coupled to the microscopic degrees of freedom of the
mirror and the internal degree of freedom of the particle. In our simple model, the plate is formed by uncoupled
unidimensional harmonic oscillators, each of them interacting locally in position with the vacuum field. The
macroscopic trajectory of the particle was externally fixed, and its internal degree of freedom was also an
unidimensional harmonic oscillator, also coupled in position with the scalar field, resulting in a dipole-like interaction.

We first studied the dissipative effects over the system. In order to do that, we followed the same approach we
had used in our previous work \cite{friction}: we calculate the in-out effective action for the whole system, which
presented an imaginary part that accounted for the non-vanishing probability of decay of the system's initial
state due to friction. We found that these dissipative effects depended on the relative velocity between the particle
and the mirror, and on the characteristic of the materials. We also found that, within our simple model,  the
imaginary part of the effective action for a particle moving in front of a mirror is the same as the imaginary part
of the effective action per unit of area for two mirrors in parallel relative motion \citep{friction}.

Then, we switched to the Schiwinger-Keldysh or CTP formalism in order to study the decoherence suffered by the
particle due to its interaction with the vacuum field fluctuations, and its effective (mediated by the vacuum field)
interaction with the mirror.  We calculated the CTP effective action describing the influence of the vacuum field and
the microscopic degrees of freedom of the mirror over the internal degree of freedom of the particle. This action of
influence allowed us to estimate the decoherence time, in which two different quantum states for the particle could
no longer be distinguished from one another. We found the decoherence time to decrease with the presence of the
mirror, and to be minimized when the characteristic frequencies of the mirror and the particle are the same (resonant case).
We also found that the relative velocity between the mirror and the plate increases the decoherence, as does the increasing of
any of the coupling constants. { We expect that in a Ramsey-like interference
experiment, the parameters of our model could be chosen in a way that would maximize the decoherence effects. As we have 
modeled the atom as the lowest energy levels of a harmonic oscillator, it is possible to choose its characteristic frequency close to the 
resonance with the plate. In addition, with a non-vanishing relative velocity, we expect decoherence effects to be observed by means of the attenuation of the contrast in the Ramsey fringes.}

\section*{Acknowledgements}
This work was supported by ANPCyT, CONICET, and UBA. FCL would like to thank ICTP - Trieste 
where part of this works has been done as a Simons Associate. MBF would like to thank LANL where part of
this works has been done as a visitor. The authors would also like to thank Diego A. R. Dalvit, C\'esar D. Fosco,
Francisco D. Mazzitelli, and Juan Pablo Paz for valuable insights and discussions.

\end{document}